\documentclass{aa} 
\usepackage{graphicx}
\usepackage{txfonts,textcomp}
\usepackage{hyperref}
\usepackage{natbib}
\usepackage{orcidlink} 
\usepackage{xcolor}

\begin{document} 

   \title{Connectivity rewiring and filament formation by active region emergence observed with Solar Orbiter}

   \author{Xiaohong Li\inst{1} \fnmsep\thanks{\hbox{\email{lixiaohong@mps.mpg.de}}}
   \orcidlink{0000-0001-8164-5633}
   \and
   Sami K. Solanki\inst{1,6} \orcidlink{0000-0002-3418-8449} 
   \and
   Thomas Wiegelmann \inst{1} \orcidlink{0000-0001-6238-0721} 
   \and
   Gherardo Valori\inst{1} \orcidlink{0000-0001-7809-0067} 
   \and 
   Daniele Calchetti\inst{1} \orcidlink{0000-0003-2755-5295} 
   \and
   Johann Hirzberger\inst{1} 
   \orcidlink{0009-0003-4886-3980}
   \and
   Clara Froment\inst{2} \orcidlink{0000-0001-5315-2890}
   \and
   Henrik Eklund\inst{3} \orcidlink{0000-0001-9597-3726}
   \and
   Nicolas Poirier\inst{2} \orcidlink{0000-0002-1814-4673}
   \and
   Luc Rouppe van der Voort\inst{4,5} \orcidlink{0000-0003-2088-028X}
   \and 
   Elias R. Udnæs \inst{4,5} \orcidlink{0009-0008-9823-7810}
   \and  
   Joachim Woch\inst{1} \orcidlink{0000-0001-5833-3738} 
   \and
   Achim Gandorfer\inst{1} \orcidlink{0000-0002-9972-9840}  
   }
   
   \institute{Max Planck Institute for Solar System Research, Göttingen D-37077, Germany 
   \and
   LPC2E, OSUC, Univ Orleans, CNRS, CNES, F-45071 Orleans, France
   \and
   European Space Agency (ESA), European Space Research and Technology Centre (ESTEC), Keplerlaan 1, 2201 AZ Noordwijk, The Netherlands
   \and
    Institute of Theoretical Astrophysics, University of Oslo, P.O. Box 1029 Blindern, N-0315 Oslo, Norway
    \and
    Rosseland Centre for Solar Physics, University of Oslo, P.O. Box 1029 Blindern, N-0315 Oslo, Norway
    \and
    School of Space Research, Kyung Hee University, Yongin, Gyeonggi 17104, Republic of Korea
   }          
              
   \date{Received ; accepted }

  \abstract
    {Using an 11-day dataset from Solar Orbiter, combining EUV imaging, H$\alpha$ observations, vector magnetograms and nonlinear force-free field extrapolations, we track the evolution from the emergence of a new active region to the reconfiguration of coronal connectivity. We find that the restructuring proceeds through a sequence of localized reconnection events that propagate along the filament channel, producing successive brightenings and bidirectional flows. This process leads to the gradual erosion of the original filament and the simultaneous formation of a new, highly sheared filament connecting the two regions. Magnetic modeling reveals a transition from a sheared arcade supporting the initial filament to a twisted flux rope underlying the newly formed structure. Our results demonstrate that flux emergence can reorganize filament channels through a spatially and temporally ordered sequence of reconnection events, rather than a single impulsive episode. This sequential reconnection front provides a natural mechanism for filament formation in evolving active regions and highlights the importance of continuously changing magnetic connectivity in the corona.}


   \maketitle
   \nolinenumbers
\section{Introduction}  \label{sec:intro}

Solar active regions (ARs) are the visible manifestations of magnetic flux emerging from the solar interior into the photosphere and above, where it concentrates into bipolar magnetic patches that can evolve into pores and sunspots and, ultimately, mature AR magnetic systems \citep{Parker1955, Fan2021}. The emergence process is not merely a surface phenomenon: as flux accumulates and reorganizes, the coupled photosphere–chromosphere–corona system develops increasingly structured magnetic connectivity that shapes the coronal structures (loops, arcades, fan systems) and governs how and where energy can be stored \citep{Cheung2014, Toriumi2019}. Because the most damaging space-weather drivers originate in magnetically complex regions, understanding how flux emergence builds active regions, particularly in their earliest, most rapidly evolving phases, is a central problem in contemporary solar physics. 

In the photosphere, the earliest emergence often appears as small, opposite-polarity concentrations that grow and separate while horizontal field becomes prominent between them, consistent with an emerging $\Omega$-loop geometry that arises as rising flux interacts with convection \citep{Fan2001, Archontis2004}. Chromospheric and transition-region signatures frequently accompany these photospheric changes. For example, arch filament systems are widely regarded as atmospheric tracers of emerging flux \citep{Bruzek1967, Bruzek1969}, and recent high-resolution/high-cadence observations have shown that small-scale brightenings within such systems can be associated with reconnection-related dynamics and may reflect the interaction of newly emerging flux with preexisting field structures \citep{Tiwari2022, Chen2024}. These early interactions are important because they can precondition the developing region by creating shear, mixed polarities, and stressed connectivity that later support energetic activity.

A mature active region can host a coronal magnetic field that departs substantially from its potential configuration, storing significant amounts of free magnetic energy in electric currents  \citep{Priest2002}. Solar eruptions are widely understood to be powered by this stored energy, which is gradually built up through the combined action of flux emergence and photospheric motions that stress the coronal field \citep{Forbes2000, Sun2012, Green2018}. In particular, flux emergence plays a central role in transporting magnetic flux, helicity, and free energy from the convection zone into the corona, where these quantities accumulate and contribute to the formation of strongly non-potential magnetic configurations, especially along polarity inversion lines (PILs) \citep{Fang2012, Georgoulis2012, Kontogiannis2025}. Numerical simulations further show that the Lorentz force associated with emerging flux naturally drives shear and rotational flows, injecting Poynting flux into the corona and promoting the formation of sheared arcades and, in some cases, flux-rope-like structures through reconnection and flux cancellation \citep{Moore2001, Aulanier2010, Wang2026}.

Beyond the buildup of non-potentiality, the interaction between emerging flux and the preexisting coronal field critically determines the subsequent evolution of the system \citep{Torok2024}. Depending on the magnetic configuration and the efficiency of reconnection with the overlying field, emerging regions may either remain confined or develop into eruptive systems \citep{Torok2005, McKevitt2026}. Observations and simulations indicate that flux emergence can destabilize preexisting coronal structures, including filaments and their overlying arcades, through magnetic interaction with the ambient field, often triggering eruptions \citep{Feynman1995, Wang1999, Chen2000, LiX2023}. At the same time, flux emergence can also stabilize coronal structures by enhancing the confining magnetic tension or reducing the upward net force \citep{Chen2000, Zhou2025a}. In this framework, flux emergence is not only a driver of energy buildup, but also a key agent of large-scale magnetic restructuring, capable of modifying coronal connectivity over extended regions. 

However, while the role of flux emergence in energizing active regions and triggering eruptions is well established, its ability to reorganize coronal magnetic connectivity in a spatially and temporally coherent manner remains less well constrained observationally. In particular, how reconnection driven by flux emergence propagates through preexisting magnetic systems, and how this process contributes to the formation, transformation, or disappearance of filament channels, is still not fully understood. Addressing these questions requires observations that can simultaneously capture long-term evolution, multilayer coupling, and the spatial progression of coronal restructuring.

The Solar Orbiter \citep{Muller2020} mission provides a unique opportunity to meet these requirements. It includes a dedicated observing plan (SOOP; \citealt{Zouganelis2020}), the Active Region Long-Term Evolution SOOP (\emph{R\_SMALL\_MRES\_MCAD\_AR-Long-Term}), designed to monitor solar active regions continuously over extended periods. This SOOP coordinates multiple remote-sensing instruments to obtain multiwavelength observations from the photosphere to the corona. The long-duration coverage enables the study of the gradual formation and evolution of active regions, including the emergence, interaction, and decay of magnetic flux \citep{Finley2025, Barczynski2025, Sinjan2025}.

For studies of magnetic flux emergence, the AR-long-term SOOP provides several key advantages. Continuous high-resolution observations allow the tracking of the magnetic and coronal evolution from the earliest stages of emergence through the development of pores, sunspots, and coronal structures. The combination of vector magnetic field measurements from the Polarimetric and Helioseismic Imager (SO/PHI; \citealt{Solanki2020}) and EUV imaging from the Extreme Ultraviolet Imager (SO/EUI; \citealt{Rochus2020}) makes it possible to investigate how newly emerging magnetic flux interacts with the preexisting coronal magnetic field and how this process contributes to energy buildup and solar eruptions. In this paper, we exploit Solar Orbiter’s high spatial and temporal resolution, together with its evolving viewing angles, to track an existing AR over 11 days with Solar Orbiter, capturing the emergence of a new AR in its vicinity. In Sect.~\ref{sec:obs} we give an overview of the observations. In Sect.~\ref{sec:res}, we present the AR evolution, connectivity changes, and filament reconfiguration. The conclusion and discussion are presented in Sect.~\ref{sec:con}.

  \begin{figure*}[!htbp]
  \centering
  \includegraphics[trim=0cm 0cm 0cm 0cm, clip, width=\textwidth] {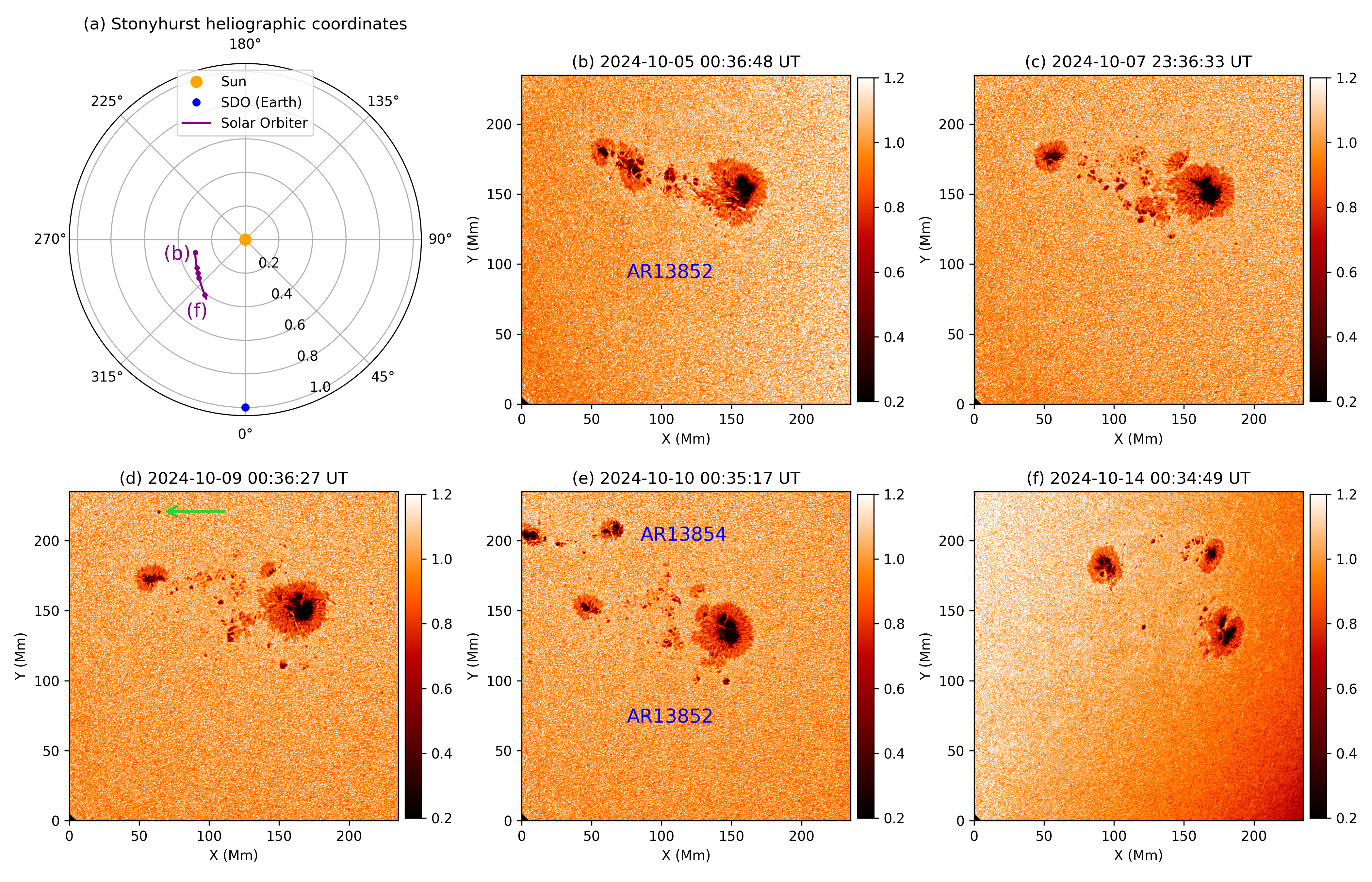}
     \caption{Relative positions of Solar Orbiter and SDO in Stonyhurst heliographic coordinates between 2024 October 5 and 15 (panel a) and SO/PHI-HRT continuum intensity maps (panels b--f) showing the evolution of the preexisting AR~13852 and the newly appearing AR~13854 during the observing interval. In panel (a), the purple curve traces the Solar Orbiter orbit, and the points mark its position at the times corresponding to the continuum-intensity maps in panels (b)--(f). The green arrow near the top of panel (d) points to the first pore associated with AR~13854.}
        \label{Fig1}
  \end{figure*}

\section{Observations}  \label{sec:obs}

The present study was based on coordinated observations from the AR-Long-Term SOOP. The target region was followed from 2024 October 5 to October 15, during which time Solar Orbiter continuously monitored one existing AR and its surroundings. Figure~\ref{Fig1}(a) shows the relative positions of Solar Orbiter and SDO during the observing campaign. Over this period, the spacecraft moved outward from about 0.3~AU to 0.4~AU, resulting in a gradual change in spatial sampling and viewing geometry, as shown in panels (b)--(f) with the corresponding orbital locations marked by the purple points in panel (a).

Because the observing campaign was carried out close to perihelion, the Solar Orbiter data provide an exceptionally detailed view of the photospheric and coronal evolution. We used observations from the SO/EUI to study the evolution of the corona. Within this SOOP, the Full Sun Imager (SO/EUI-FSI) supplies full-disk context images with a cadence of 10~min, while the High Resolution Imager (SO/EUI-HRI) acquires high-resolution EUV observations during a dedicated one-hour observing window per day. The restriction to one such window per day was maintained to satisfy telemetry constraints. We employed the level-2 SO/EUI-HRI 174~{\AA} data to examine the coronal fine structure of the active region at high spatial resolution. The SO/EUI-HRI has a field-of-view (FOV) of 2048 × 2048 pixels and a plate scale of 0\farcs492~pix$^{-1}$, which corresponds to a linear scale of approximately $\sim$ 107~km~pix$^{-1}$ at 0.3~AU and $\sim$ 147~km~pix$^{-1}$ at 0.4~AU at disk center, enabling us to resolve the small-scale evolution of coronal loops and localized brightenings occurring in the target region. In addition, we used SO/EUI-FSI 174~{\AA} and 304~{\AA} observations to place the high-resolution SO/EUI-HRI data into a broader full-Sun context and to follow the large-scale coronal and transition-region evolution throughout the full observing interval.

To investigate the magnetic evolution at the photospheric level, we used measurements from the High Resolution Telescope (HRT; \citealt{Gandorfer2018}) of the SO/PHI. The instrument records the full Stokes profiles of the Fe I 6173~{\AA} spectral line at five positions across the line and one nearby continuum wavelength. These data allow us to derive continuum intensity, line-of-sight (LOS) velocity, and vector magnetic-field information. During this SOOP, SO/PHI-HRT was operated with a cadence of 1~h, with an interruption between 00:00 UT and 14:30 UT on October 8. For the present analysis, we used the level-2 magnetic and intensity products provided by the SO/PHI pipeline. The reduction and calibration of the SO/PHI-HRT observations follow the procedures described by \citet{Kahil2022}, \citet{Sinjan2023}, and \citet{Bailen2024}. The spatial sampling of SO/PHI-HRT is about 0\farcs5~pix$^{-1}$, corresponding to $\sim$ 110~km~pix$^{-1}$ at 0.3~AU and $\sim$ 147~km~pix$^{-1}$ at 0.4~AU (disk center).

To complement the Solar Orbiter observations with an Earth-based viewpoint, we also used data from the Atmospheric Imaging Assembly (AIA; \citealt{Lemen2012}) on board the Solar Dynamics Observatory (SDO; \citealt{Pesnell2012}). SDO/AIA provides full-disk EUV and UV imaging with a plate scale of 0\farcs6~pix$^{-1}$ (about 435~km at the Sun). In this work, we primarily employed the SDO/AIA 304~{\AA} channel, which has a peak response at log T $\sim$ 4.7 and a cadence of 12~s. Because Solar Orbiter and SDO observe the Sun from different heliocentric distances and vantage points, the two datasets are not directly comparable in time without correction. We therefore shifted the Solar Orbiter observation times to the corresponding Earth-referenced times by accounting for the difference in light-travel time between the two spacecraft. This correction was applied throughout the analysis. 

We also used a subset of observations from the Swedish 1-m Solar Telescope \citep[SST;][]{Scharmer2003} on La Palma. Imaging spectroscopy in the H$\alpha$ line was performed using the CRISP instrument \citep{Scharmer2008}, which sampled H$\alpha$ at 15 line positions within $\pm$2.8~{\AA} around the line core in about 6~s. The spatial sampling was 0\farcs044~pix$^{-1}$, corresponding to approximately 32~km~pix$^{-1}$ at disk center. We note that this value represents the detector plate scale rather than the effective spatial resolution. The diffraction-limited Rayleigh resolution of the SST at H$\alpha$ is $1.22\lambda/D \approx 0\farcs16$, corresponding to about 120~km on the Sun, and reaching this resolution requires sufficiently good seeing conditions together with image restoration. The data were processed following the SSTRED data reduction pipeline \citep{2015A&A...573A..40D, 2021A&A...653A..68L}, which includes multi-object multi-frame blind deconvolution (MOMFBD) image restoration \citep{2005SoPh..228..191V}. To cover the full extent of the target AR, a mosaic observing strategy was employed. The telescope was moved over an overlapping grid of 4 $\times$ 3 or 5 $\times$ 4 pointings to cover the full AR. The adjacent pointings were co-aligned and combined to produce a large-FOV map of the AR.

 \begin{figure*}[!htbp]
  \centering
   \includegraphics[trim=0cm 0cm 0cm 0cm, clip, width=0.9\textwidth]{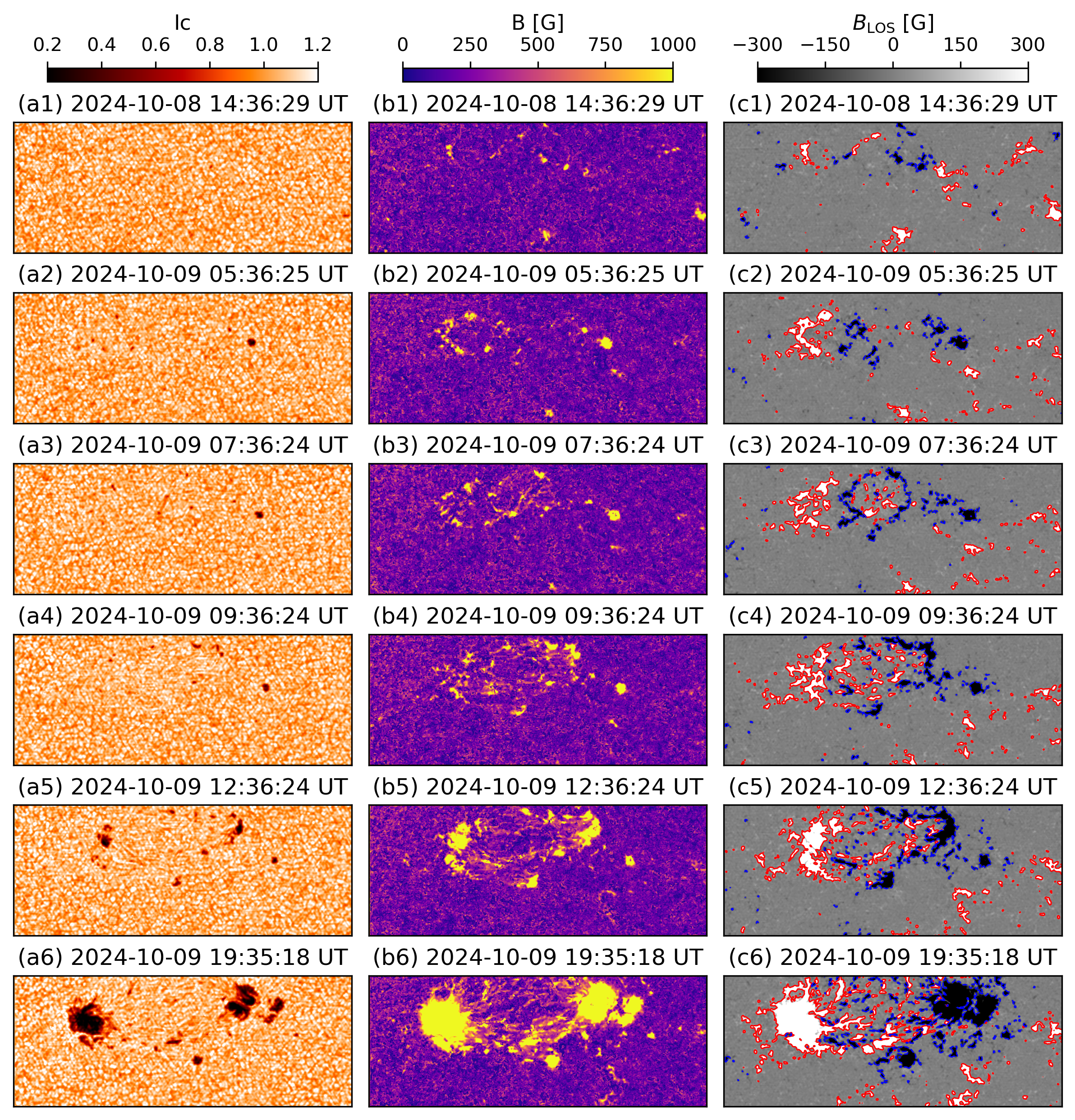}
     \caption{SO/PHI-HRT continuum intensity (panels a1--a6), magnetic field strength (panels b1--b6) and the LOS magnetic field (panels c1--c6) maps showing the formation of AR~13854. Red and blue contours in panels (c1)--(c6) represent the positive and negative polarities at ± 100G, respectively.
    }
        \label{Fig2}
  \end{figure*}

\section{Analysis and results} \label{sec:res}

Starting from 2024 October 4 at 21:31 UT, Solar Orbiter observed one AR near the solar east limb, which was later numbered NOAA AR~13852 after it became visible from the Earth. The region initially exhibited a complex morphology composed of multiple sunspot fragments, as shown in Fig.~\ref{Fig1}(b). At this stage, AR~13852 was already in a continuing emergence and organization phase, with ongoing magnetic flux supply from below the photosphere. During this phase, patches of the same magnetic polarity progressively coalesced, while opposite-polarity elements underwent mutual cancellation along the PIL \citep{Valori2012}. As a result, the magnetic configuration evolved toward a more organized bipolar structure (see panel c). Beginning on October 9, additional small-scale magnetic and intensity features started to appear near the upper left corner of the SO/PHI-HRT FOV, as indicated by the green arrow in panel (d). These features mark the onset of a new emerging active region, subsequently designated as AR~13854, located to the northeast of AR~13852 (see panel e). Over the following days, AR~13854 exhibited sustained growth in both size and magnetic complexity. Meanwhile, AR~13852 entered a gradual decay phase: from panel (d) to (f), both sunspots became smaller, and the eastern sunspot decayed into a pore and nearly disappeared toward the end of the observing period.

\subsection{Emergence of the new AR}
Figure~\ref{Fig2} illustrates the early formation stage of AR~13854. In the continuum intensity maps (left column), the emergence of magnetic flux is manifested by the intermittent appearance of pores. These features are initially faint and highly fragmented, but progressively expand and merge as magnetic flux accumulates. The evolution is more clearly revealed in the magnetic field diagnostics shown in the middle and right columns. In the magnetic field strength maps (middle column), numerous small-scale magnetic concentrations appear sequentially across the region. At early times, these elements are weak and spatially dispersed. As flux emergence proceeds, same-polarity concentrations migrate and coalesce, forming larger magnetic structures, often connected by elongated, filamentary features (often also visible in the continuum intensity). The LOS magnetic field (B$_{\mathrm{LOS}}$; right column) further reveals that the emerging flux is organized into multiple small-scale bipolar pairs. These bipoles do not emerge simultaneously, but rather appear sequentially over time. Initially, they are loosely distributed and only weakly connected. As the evolution continues, opposite polarities within individual bipoles separate, while magnetic elements of the same polarity from neighboring bipoles migrate and merge, leading to the formation of increasingly coherent, large-scale bipolar structures. This process ultimately results in the development of pores and the subsequent formation of sunspots. The different magnetic behaviors of AR~13852 and AR~13854 therefore reflect their different evolutionary stages: AR~13852 was already undergoing organization and subsequent decay, whereas AR~13854 was captured during its initial flux-emergence phase.

\subsection{Connectivity rewiring}
\begin{figure*}[!htbp]
  \centering
  \includegraphics[width=0.8\textwidth]{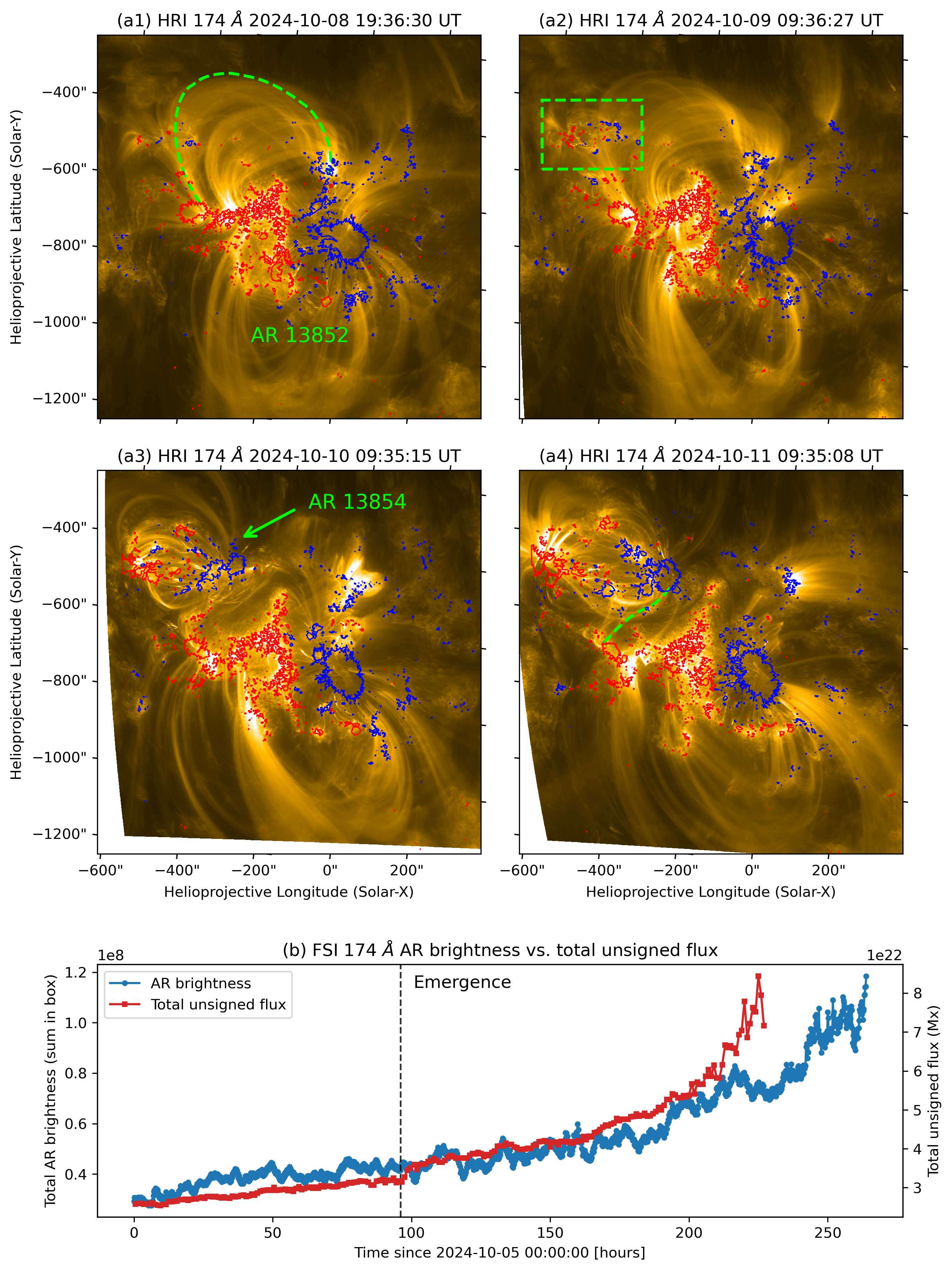}
     \caption{Panels (a1)--(a4): SO/EUI-HRI 174~{\AA} images showing the connectivity change of the coronal loops caused by the emergence of AR~13854. Red and blue contours represent the positive and negative polarities at ± 500G, respectively. The dashed curve in each of panels (a1) and (a4) denotes a selected field line, and the dashed box in panel (a2) indicates the newly emerged AR. Panel (b): Changes in 174~{\AA} total intensity (blue curve) and total unsigned flux (red curve) between 2024 October 5 and 15 within the whole FOV shown in the upper panels. High-cadence SO/EUI-HRI 174~{\AA} movies within each individual observing window are publicly available via JHelioviewer or the EUI data archive.}
    \label{Fig3}
  \end{figure*}

In the corona, the evolution of the emerging magnetic field was traced by the loop structures observed in the 174~{\AA} channel, as shown in Figure~\ref{Fig3}. On 2024 October 8, AR~13852 exhibits a relatively simple bipolar configuration, with coronal loops connecting the dominant positive and negative polarities. The corona overlying the emergence site of AR~13854 is initially faint and diffuse, with no clearly discernible loop structures in the 174~{\AA} channel, indicating that a bright and organized 174~{\AA} loop system had not yet developed at this stage. As the magnetic flux continues to emerge and concentrate in the photosphere, small coronal loops begin to appear intermittently (see the dashed box in panel a2). With further flux accumulation and organization, these loops become progressively brighter and more coherent, forming a system of low-lying coronal loops rooted in the emerging bipoles of AR~13854, as shown in panel (a3). This transition indicates an increase in the observed 174~{\AA} emission and the development of more organized low-lying magnetic connectivity, possibly reflecting enhanced plasma filling and heating along the newly emerged field as the magnetic structure becomes stronger and more coherent. 

The emerging loops of AR~13854 expand upward and interact with the preexisting loops of AR~13852, leading to the formation of loops that connect the two active regions (see the dashed curve in panel a4). Meanwhile, the northern loops of AR~13852 fade from view, consistent with a reconfiguration of magnetic connectivity. As the evolution proceeds, the overall coronal morphology becomes increasingly complex, reflecting the continuous injection of magnetic flux and the associated restructuring of the coronal field. 

To quantify the relationship between coronal emission and photospheric magnetic evolution, we computed the total unsigned magnetic flux and the EUV intensity within the region of interest encompassing the two ARs. The temporal evolution of these quantities is shown in panel (b). The unsigned magnetic flux increases rapidly during the emergence phase, reflecting the continuous injection of magnetic field into the photosphere. The EUV intensity exhibits a broadly similar increasing trend, but with a systematic delay relative to the flux evolution. This delay likely reflects the timescale required for the newly emerged magnetic field to expand into the corona and for the coronal plasma to respond, including density build-up, thermal evolution such as plasma heating through current dissipation or reconnection and subsequent cooling from higher temperatures into the 174~{\AA} response range \citep{Warren2002, Ugarte2006, Ugarte2009, Viall2011, Viall2017}.

    \begin{figure*}[!htbp]
  \centering
  \includegraphics[trim=0.4cm 0.4cm 0cm 0.5cm, clip, width=\textwidth]{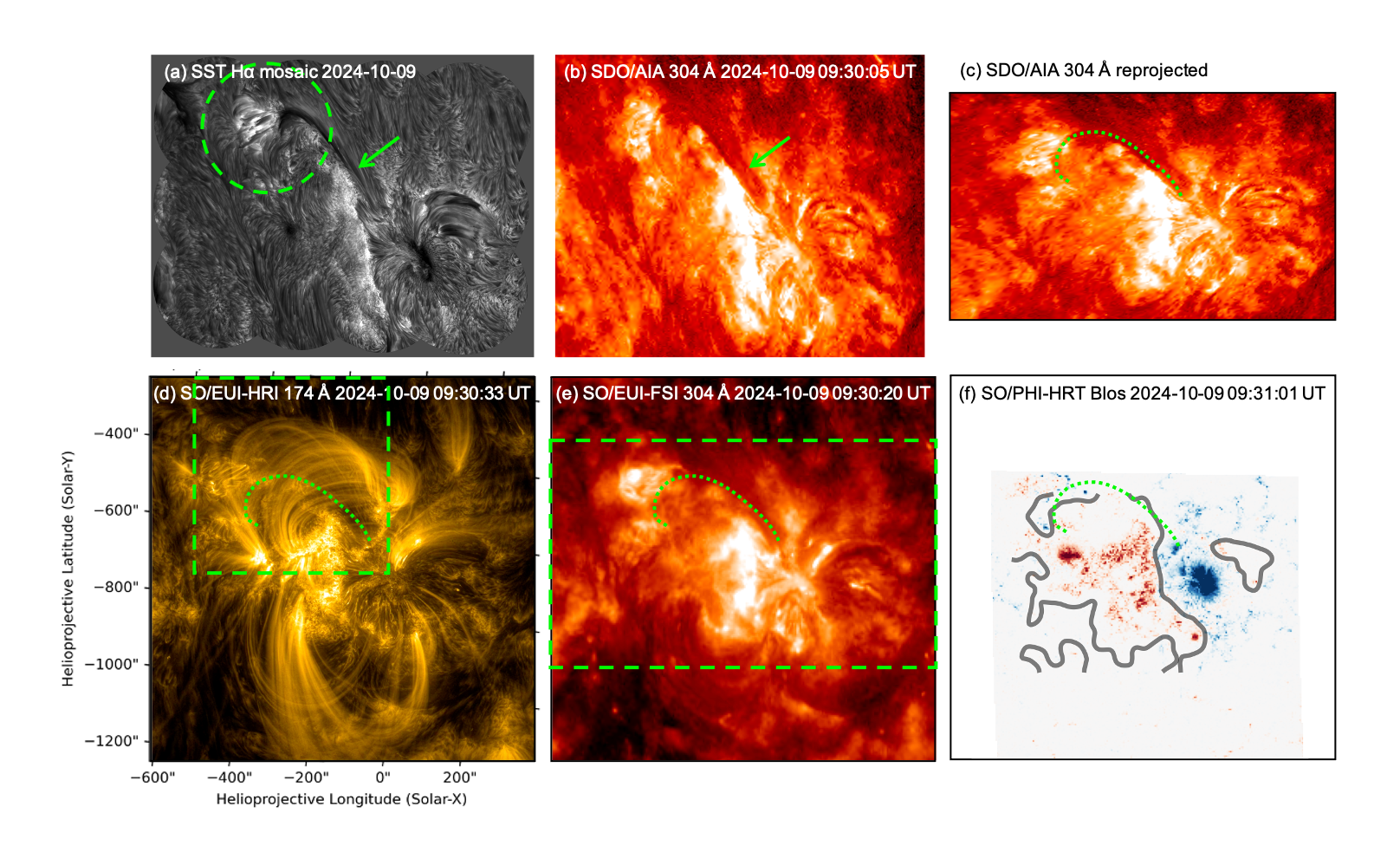}
     \caption{SST H$\alpha$ mosaic image, SDO/AIA 304~{\AA} image, SDO/AIA 304~{\AA} image reprojected to the Solar Orbiter view, SO/EUI-HRI 174~{\AA} image, SO/EUI-FSI 304~{\AA} image, and SO/PHI-HRT $B_{\rm LOS}$ map showing the filament in AR~13852. The green arrows in panels (a)--(b) and the dotted curves in panels (c)--(f) mark the filament located along the PIL of AR~13852. The circle in panel (a) outlines the FOV of Fig.~\ref{Fig5}, the box in panel (d) indicates the FOV of Fig.~\ref{Fig6}, and the rectangle in panel (e) denotes the FOV of panels (a)--(c). In panel (f), the solid black curves denote the PILs, defined as the $B_{\rm LOS}=0$ contour after Gaussian smoothing with a spatial scale of 0.5~Mm to suppress small-scale noise, and only contour segments longer than 8~Mm are retained.
     }
    \label{Fig4}
  \end{figure*}   
  
\subsection{Filament reformation}

 Prior to the emergence of AR~13854, a preexisting filament was present along the northern segment of the PIL of AR~13852. From 2024 October 9 onward, this filament was also visible from the Earth-viewing perspective, as shown in Fig.~\ref{Fig4}. In the SST H$\alpha$ mosaic and the SDO/AIA 304~{\AA} image, the filament is located close to the solar limb. Consequently, its apparent width is reduced by projection effects and the structure appears narrow and elongated in panels (a) and (b). To compare the Earth-view observations with those from Solar Orbiter, we reprojected the SDO/AIA 304~{\AA} image to the Solar Orbiter viewing geometry. The reprojected AIA image in panel (c) shows good morphological agreement with the SO/EUI-FSI 304~{\AA} image in panel (e), confirming that the same filament structure is observed from the two viewpoints. The SO/PHI-HRT LOS magnetogram further shows that the filament lies along the PIL separating opposite magnetic polarities in AR~13852. To identify the PILs, the $B_{\rm LOS}$ map was smoothed with a Gaussian kernel of 0.5~Mm to suppress pixel-scale noise, and the resulting $B_{\rm LOS}=0$ contour was extracted, retaining only segments longer than 8~Mm (see panel f). The filament spine follows the PILs, while its two ends are rooted in magnetic regions of opposite polarity: the eastern (left) footpoint is anchored in a positive-polarity region, whereas the western (right) footpoint is rooted in a negative-polarity region. This magnetic configuration is consistent with a filament channel supported by sheared or dipped magnetic field lines above the PIL.
 
\begin{figure*}[!htbp]
  \centering
  \includegraphics[trim=0.6cm 1cm 0.4cm 1cm, clip, width=0.95\textwidth]{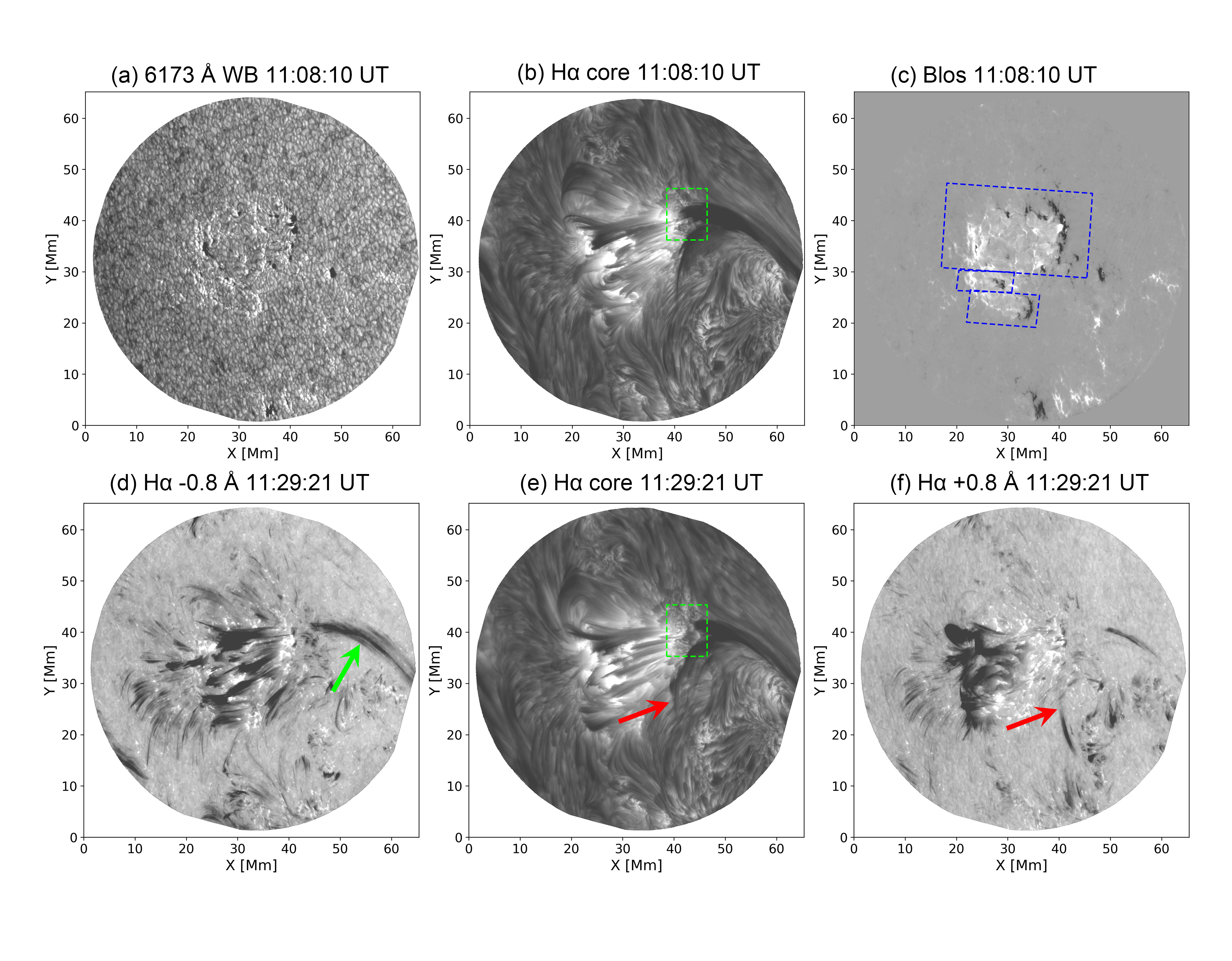}
     \caption{SST observations of the interaction between the preexisting filament and the emerging flux in AR~13854 on 2024 October 9. Panel (a) displays the CRISP Fe~I 6173~{\AA} wide-band image, panel (b) shows the H$\alpha$ line-core image, and panel (c) shows the corresponding LOS magnetic field map. Panels (d)--(f) show the H$\alpha$ images obtained at a later time at H$\alpha-0.8$~{\AA}, H$\alpha$ line center, and H$\alpha+0.8$~{\AA}, respectively. The green boxes in panels (b) and (e) mark the region where the filament interacts with the emerging magnetic flux. The blue boxes in panel (c) outline three emerging bipoles. The red arrows indicate downflowing filament material, which is visible in the red-wing H$\alpha+0.8$~{\AA} image, while the green arrow indicates upward-moving filament material, which is visible in the blue-wing H$\alpha-0.8$~{\AA} image. The associated movie (Movie1.mp4) is available online.
     }
    \label{Fig5}  
  \end{figure*}

 As the new active region, AR~13854, starts to emerge, the magnetic environment near the eastern (left) footpoint of the preexisting filament is progressively disturbed. The activation of the filament during this phase is captured by the high-resolution SST observations shown in Fig.~\ref{Fig5}. The field of view of Fig.~\ref{Fig5} corresponds to the circled region in Fig.~\ref{Fig4}(a), covering both the emerging-flux region and the eastern part of the filament. The CRISP Fe~I 6173~{\AA} wide-band image in panel (a) reveals small-scale photospheric changes associated with flux emergence, including the development of dark pores and disturbed granulation in the emerging region. The corresponding LOS magnetogram in panel (c) shows several emerging bipoles, marked by the blue boxes. The opposite-polarity patches within these bipoles separate and strengthen with time, indicating the continuous injection of new magnetic flux through the photosphere. The co-spatial H$\alpha$ line-core image in panel (b) shows compact brightenings and evolving dark fibril-like structures around the emerging polarities, which are typical chromospheric signatures of an emerging flux region interacting with the ambient magnetic field. Such localized chromospheric brightenings are commonly found in the vicinity of strong LOS magnetic-field concentrations \citep{Eklund2020,Eklund2023}. 
 
 As the emerging bipoles expand toward the preexisting filament channel, the filament becomes activated and appears to split into two branches near the interaction region. The subsequent H$\alpha$ wing and line-core images in panels (d)--(f), together with the accompanying movie (Movie1.mp4), reveal the motion of filament material during this interaction. The green boxes in panels (b) and (e) mark the interface between the emerging flux and the filament. Comparison between the earlier H$\alpha$ line-core image in panel (b) and the later one in panel (e) shows that the preexisting filament structure weakens near the interaction interface, while more dark material appears to accumulate around the emerging footpoints. The northern branch shows enhanced signatures in the H$\alpha$ blue-wing image, indicating upward or blueshifted motion of filament plasma, as marked by the green arrow in panel (d). In contrast, the southern branch shows enhanced signatures in the H$\alpha$ red-wing image near the footpoints, indicating redshifted downflows of filament material, as marked by the red arrow in panel (f). These observations suggest that the newly emerged magnetic flux interacts with the preexisting filament-channel field, leading to the partial disruption and redistribution of filament material.

\begin{figure*}[!htbp]
  \centering
  \includegraphics[trim=0cm 0cm 0cm 0cm, clip, width=\textwidth]{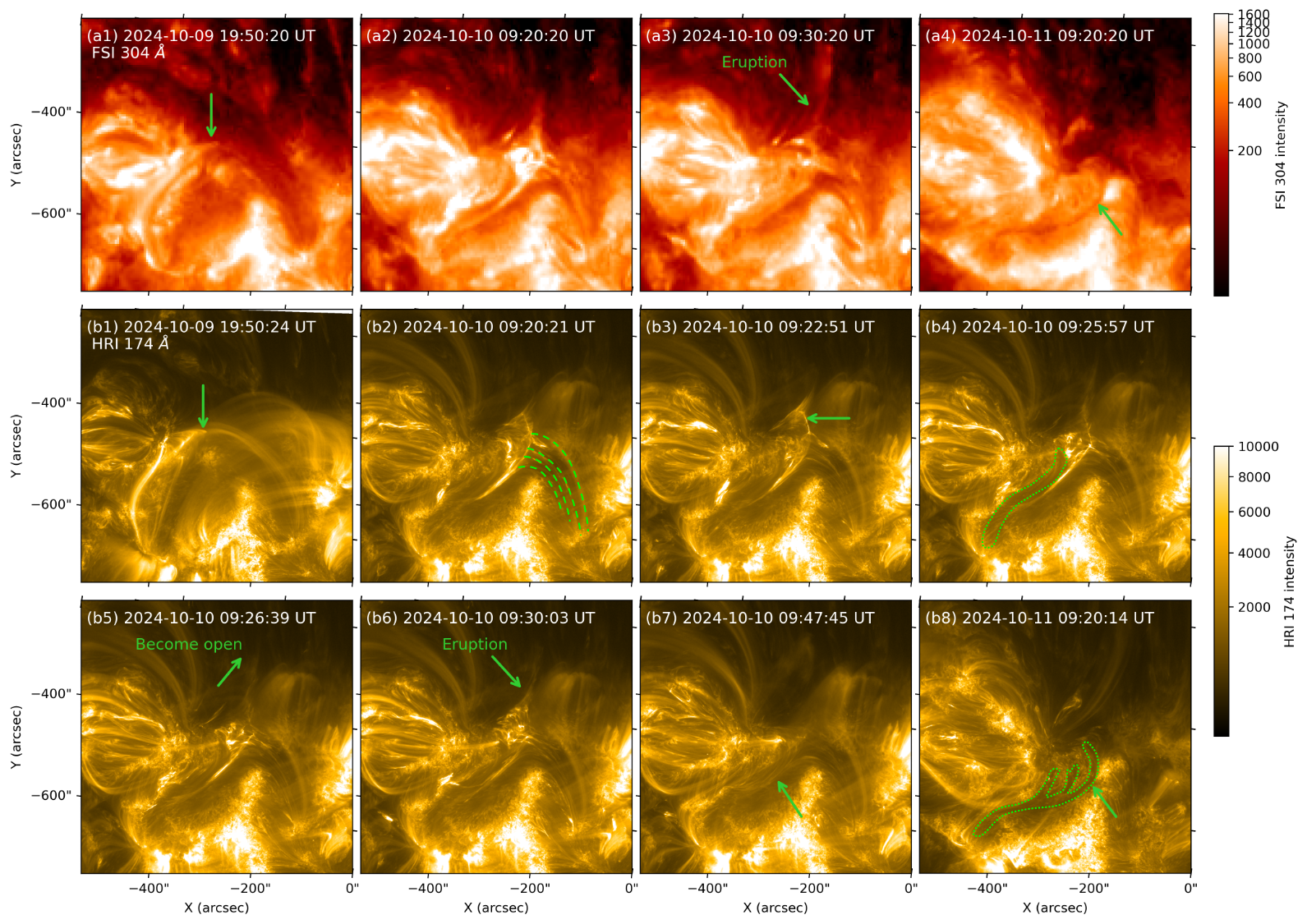}
     \caption{SO/EUI-FSI 304~{\AA} images (panels a1--a4) and SO/EUI-HRI 174~{\AA} images (panels b1--b8) showing the partial eruption and indicating the magnetic reconfiguration of the filament. The green arrows in panels (a1) and (b1) indicate the apparent X-shaped site where the new emerging loops interact with the filament. The dashed curves in panel (b2) denote the fine structures in the filament. The arrow in panel (b3) points to the current sheet. The arrows in panels (a4), (b7) and (b8) mark the newly formed filament.}
        \label{Fig6}
  \end{figure*}

 The subsequent large-scale response of the filament system is captured by the SO/EUI observations shown in Fig.~\ref{Fig6}. While the SST H$\alpha$ observations in Fig.~\ref{Fig5} reveal the low-atmosphere activation and initial splitting of the filament near the emerging bipoles, the SO/EUI 304~{\AA} and 174~{\AA} images trace the continuation of this process at larger spatial scales. The SO/EUI-FSI 304~{\AA} images in panels (a1)--(a4) show the evolution of the cool filament material, while the SO/EUI-HRI 174~{\AA} images in panels (b1)--(b8) reveal the associated coronal brightenings and fine loop structures. At the early stage, panels (a1) and (b1) show that an apparent X-shaped interaction site is visible between the newly emerging loops and the preexisting filament, as indicated by the green arrows. This feature is consistent with the interaction region identified in the SST H$\alpha$ observations but much higher, suggesting that the newly emerged magnetic field continues to reconnect with the filament-channel field above the low atmosphere. Bright coronal loops connecting AR~13852 and AR~13854 appear on the left of this site, indicating that the magnetic connectivity between the two active regions is being reorganized. As the reconnection proceeds, the original filament becomes increasingly fragmented. The SO/EUI-HRI 174~{\AA} image in panel (b2) resolves several fine filamentary threads, outlined by the dashed curves, indicating that the filament exhibits a multi-threaded structure. 
 
 During the following stage, the localized X-shaped feature develops into an elongated current-sheet-like structure, as indicated in panel (b3), consistent with continued magnetic reconnection between the emerging loops and part of the filament threads. Following this process, a new filament channel forms between the two active regions (outlined by the green contour in panel b4), while portions of the original filament are ejected along large-scale, outward-expanding field lines (relatively ``open'' to local structure). In panels (a3) and (b6), filament material moves outward along large-scale expanding field lines. This process repeats over time: the northern branch undergoes successive eruptions (see panel b6), whereas plasma from the southern branch is incorporated into the newly formed inter-AR filament. As a result, the original filament is progressively removed, and a new filament system connecting the two ARs is formed (see the green arrows in panels b7 and b8). In 304~{\AA} observations shown in panels (a4), the newly formed filament is clearly visible. The apparent evolution from left to right suggests that the original filament material is gradually transferred and reorganized into the inter-AR filament channel. This inter-AR filament exhibits multiple northern extensions (see the contours in panel b8), likely reflecting successive reconnection episodes at different locations along the channel.  

\begin{figure*}[!htbp]
  \centering
  \includegraphics[trim=3cm 21cm 3cm 15cm, clip, width=0.9\textwidth]{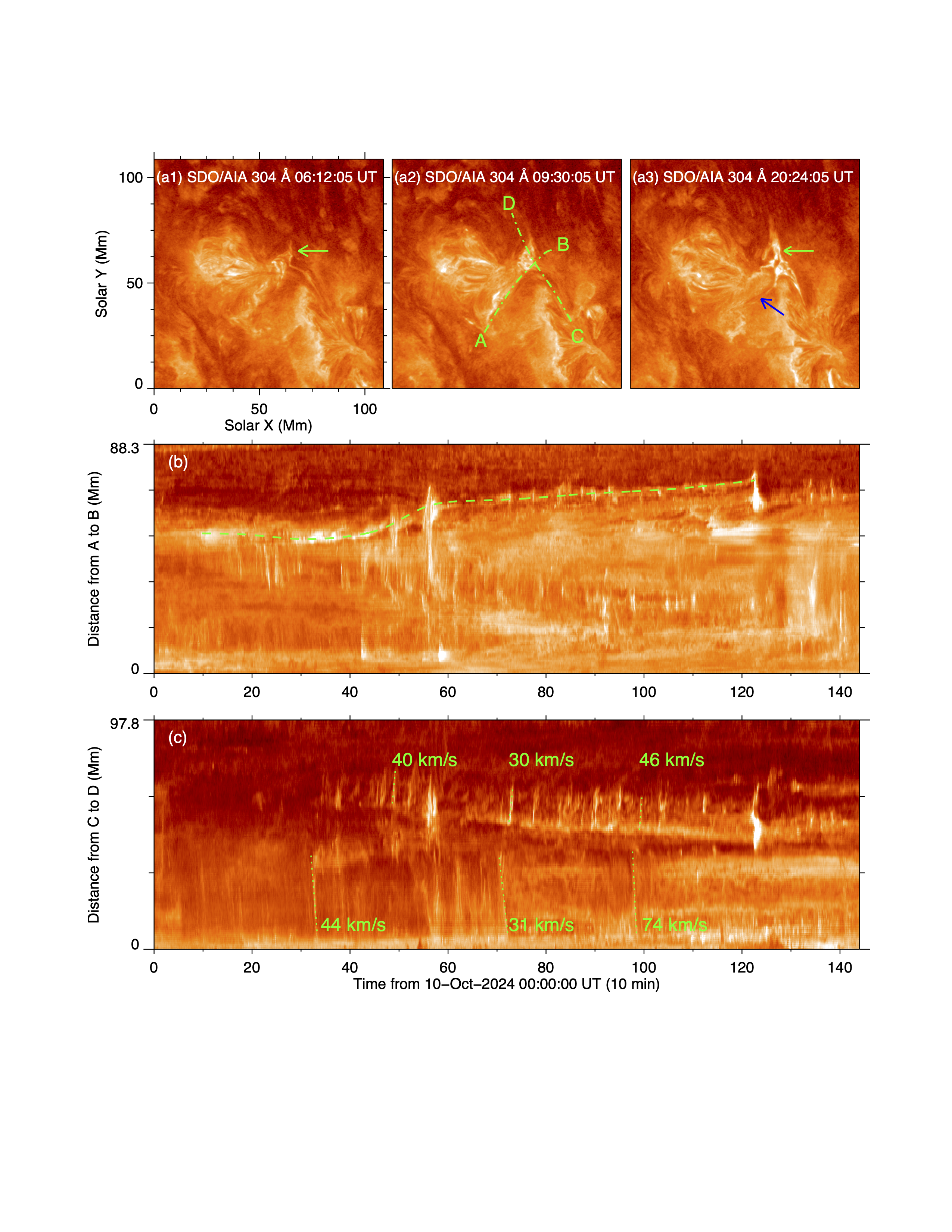}
     \caption{Panels (a1)--(a3): SDO/AIA 304~{\AA} images displaying the partial eruption and magnetic reconfiguration of the filament on 2024 October 10. The green arrows indicate the eruption of the northern branch, and the blue arrows point to the newly formed filament. The green curves in panel (a2) indicate the locations of the slits used to construct the time--distance maps shown in panels (b) and (c). Panel (b): Time--distance map along slice ``A-B'' showing the eruption of the filament. The dashed green curve denotes the migration of the brightening. Panel (c): Time--distance diagram along slice ``C-D.'' The dotted green lines denote selected features we tracked, with the corresponding projected velocities indicated. The associated movie (Movie2.mp4) is available online.}
        \label{Fig7}
  \end{figure*}

 To trace the temporal progression of the repeated reconnection, we further analyzed the filament dynamics using 304~{\AA} data from SDO/AIA. Figure~\ref{Fig7} and the accompanying movie (Movie2.mp4) show the evolution of the filament on 2024 October 10. The AIA 304~{\AA} images in panels (a1)--(a3) reveal a sequence of localized eruptions and brightenings along the filament channel. The activity starts from the left part of the filament and subsequently progresses toward the right, indicating that the magnetic reconfiguration does not occur simultaneously along the entire filament, but proceeds successively along the channel on a timescale determined by the photospheric evolution. The newly formed filament, marked by the blue arrow, becomes progressively more evident as the original filament material is redistributed. The fan-shaped brightening pattern visible in panel (a3) suggests that the reconnection geometry expands laterally as the interaction proceeds. To quantify this apparent propagation, we constructed time--distance maps along the two slits marked in panel (a2). Along slit ``A-B,'' shown in panel (b), the brightening front exhibits a clear migration toward point B, as indicated by the dashed green curve. This systematic motion is consistent with a reconnection site that propagates from left to right along the filament channel. At the same time, dark filament material is seen to move toward point A, suggesting drainage or redistribution of cool plasma following each reconnection episode. Along slit ``C-D,'' shown in panel (c), clear bidirectional flows are detected near the intersection of the two slits, suggesting a reconnection outflow region. These flows have projected velocities of the order of several tens of km~s$^{-1}$. Together, the imaging sequence and the time--distance analysis support a scenario of repeated reconnection, in which the reconnection region migrates along the filament channel and progressively transfers material into the newly formed filament.

\subsection{Coronal magnetic field extrapolation}

To investigate the evolution of magnetic connectivity associated with flux emergence, we carried out coronal magnetic field extrapolations using photospheric vector magnetograms as the lower boundary. Owing to the limited FOV of the SO/PHI-HRT observations, we constructed an extended and approximately flux-balanced boundary by combining the high-resolution SO/PHI-HRT measurements with co-temporal vector magnetograms from SDO/HMI SHARP data \citep{Bobra2014}. Prior to the extrapolation, the azimuthal ambiguity of the transverse magnetic field was resolved using the minimum-energy method \citep{Metcalf1994, Metcalf2006}. The SO/PHI-HRT inversion outputs (field strength, inclination, and azimuth) were first converted into the observer-frame components and then transformed into the local heliographic components $(B_r, B_\theta, B_\phi)$ using the formalism of \citet{Gary1990}, taking into account the spacecraft viewing geometry. The resulting vector components were then remapped onto a Cylindrical Equal-Area (CEA) grid consistent with the SHARP data products to ensure a uniform geometry and pixel scale. Within the SO/PHI-HRT FOV, the reprocessed SO/PHI-HRT data were retained, while the surrounding regions were supplemented with SHARP data, thereby providing a spatially extended boundary condition suitable for coronal magnetic field extrapolation. To validate this data-combination procedure, we performed potential-field extrapolations using SHARP data alone (2024 October 8--18) and using the combined HRT+SHARP boundary (2024 October 4--15). The two approaches yield broadly consistent large-scale magnetic topology (see Appendix~\ref{app:sharp_comparison}), confirming that the combination procedure does not introduce significant artifacts. This suggests that combining SO/PHI-HRT and SDO/HMI observations offers a feasible approach for future studies seeking to track the long-term magnetic evolution of AR beyond the limited FOV and cadence of a single instrument alone.

We performed both potential field extrapolations and nonlinear force-free field (NLFFF) extrapolations. To reduce the influence of noise-dominated weak-field regions, we applied a threshold mask to the photospheric magnetic field, excluding pixels with a field strength below 300 G from contributing to the boundary constraints. The NLFFF extrapolation was carried out using the optimization method \citep{Wheatland2000, Wiegelmann2004, Wiegelmann2012a}. This approach determines the three-dimensional (3D) magnetic field by minimizing a functional that measures the deviation from the force-free and divergence-free conditions throughout the computational domain, as well as deviations of the modeled field from the observed magnetic field at the lower boundary. In this framework, the magnetic field is iteratively adjusted to reduce the Lorentz force and the magnetic field divergence, subject to the observed photospheric boundary. Compared to potential field extrapolations, the NLFFF solutions incorporate field-aligned electric currents and are therefore capable of capturing magnetic shear and twist in active regions. The computations were performed in a Cartesian domain with uniform grid spacing, and we verified that the large-scale magnetic connectivity is consistent between the potential and NLFFF models. However, the latter exhibits enhanced non-potentiality in regions of strong magnetic gradients.

\begin{figure*}[!htbp]
  \centering
  \includegraphics[trim=0.4cm 0.2cm 0.4cm 0.5cm, clip, width=\textwidth]{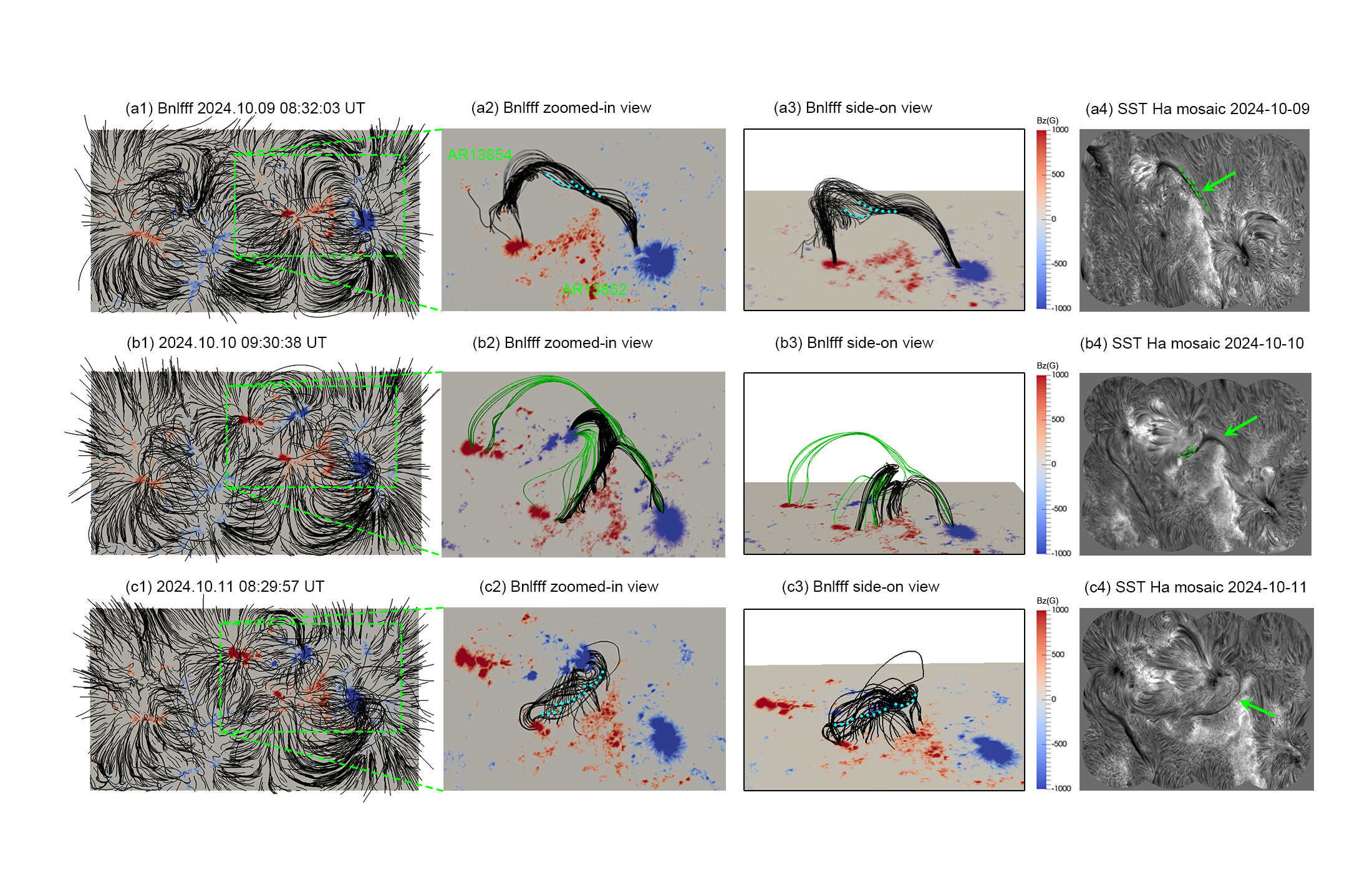}
     \caption{Magnetic restructuring of the filament system from 2024 October 9--11 in NLFFF extrapolations and SST H$\alpha$ observations. Left column: Large-scale extrapolated magnetic field. Second column: Zoomed-in top-down views of the filament-channel magnetic structure. Third column: Corresponding side-on views. Right column: Corresponding SST H$\alpha$ mosaic images. In the zoomed-in and side-on panels, black field lines trace the magnetic structure associated with the filament, whereas green curves represent the newly formed structure after reconnection. Dotted cyan curves represent magnetic dips where filament material may be supported. The green arrow in panel (a4) indicates the original filament in AR~13852. Dashed green curves in panels (a4) and (b4) represent selected threads in the filament. The green arrow in panel (c4) denotes the newly formed filament between the two ARs.}
        \label{Fig8}        
  \end{figure*}

Figure~\ref{Fig8} combines the NLFFF extrapolation results with the SST H$\alpha$ observations, showing the magnetic restructuring of the filament system from 2024 October 9 to 11. On October 9, before the main restructuring, AR~13852 shows a relatively simple bipolar magnetic configuration, with arcade-like field lines connecting the main positive and negative polarities (see panel a1). In addition, some field lines connect to the surrounding plage regions. The emerging flux region, at the left corner of the dashed rectangle, can be characterized by small-scale low-lying loops. There are also loops connecting the negative polarity of AR~13854 and the positive sunspot of AR~13852, showing the beginning of the restructuring. The preexisting filament in AR~13852 is supported by a sheared-arcade-like magnetic structure, as shown by the black field lines in panels (a2) and (a3). These field lines are inclined with respect to the local potential arcade and are more nearly aligned with the PIL, indicating enhanced magnetic shear along the filament channel. The side-on view in panel (a3) further reveals shallow dipped or nearly horizontal segments (indicated by the dotted cyan curves) along the filament spine, consistent with the expected support sites for filament material. The corresponding SST H$\alpha$ image in panel (a4) shows the original filament, marked by the green arrow. The selected filament threads, indicated by the dashed green curves, are oriented from the upper-left to the lower-right, in agreement with the direction of the sheared magnetic field in the NLFFF model. As mentioned in Fig.~\ref{Fig4}, the filament appears narrower in the SST observations due to the projection effect, and the reprojected filament morphology in Fig.~\ref{Fig4}(c) agrees even better with the extrapolated field.

On October 10, the continued emergence of AR~13854 reconnects with the eastern part of the preexisting filament channel. The NLFFF extrapolation shows that the magnetic field near the interaction region becomes more complex. Part of the original filament-supporting field, shown by the black curves in panels (b2) and (b3), is disrupted, while new magnetic connections, shown by the green curves, continue to form between AR~13854 and AR~13852. The side-on view in panel (b3) shows that part of the original dipped field lines has been removed, consistent with the partial disruption of the filament-supporting structure. Some of the filament material can be redistributed into the newly forming magnetic channel due to reconnection. This transition is consistent with the SST H$\alpha$ image in panel (b4), where the filament no longer appears as a single coherent structure. Instead, the left portion of the original filament splits into different branches, and the newly organized southern branch shows shorter threads with a different orientation, as indicated by the dashed green curves in panel (b4). Originally, the filament had long threads oriented from the upper left to the lower right. The southern part subsequently exhibited short threads oriented from the lower left to the upper right, indicating a change in the underlying magnetic connectivity.

By October 11, the magnetic topology has been substantially reconfigured. The NLFFF extrapolation reveals a newly formed inter-AR filament channel connecting AR~13854 and AR~13852, shown by the black field lines in panels (c2) and (c3). Compared with the original sheared arcade in AR~13852, this new structure appears more twisted and contains dipped segments along the field lines, as highlighted by the dotted cyan curves. These dips provide favorable locations for the accumulation of cool filament material. The corresponding SST H$\alpha$ mosaic in panel (c4) shows the newly formed filament between the two active regions, marked by the green arrow. Compared to panel (a4), the newly formed filament is much thinner, suggesting a more confined and possibly more sheared magnetic configuration. Its thread orientation is different from that of the original filament in AR~13852, consistent with the formation of a new magnetic channel through reconnection. The agreement between the modeled dipped field lines and the observed H$\alpha$ filament threads supports a scenario in which sustained flux emergence and reconnection progressively remove part of the original filament system and reorganize the remaining material into a new inter-AR filament. After the formation of this filament, the two-AR system becomes more dynamic. The newly formed filament becomes increasingly unstable and subsequently starts to erupt, accompanied by enhanced flare activity, including more than ten flares stronger than C class between October 11 and 15.

\begin{figure*}[!htbp]
  \centering
  \includegraphics[trim=0cm 0cm 0cm 0cm, clip, width=0.9\textwidth]{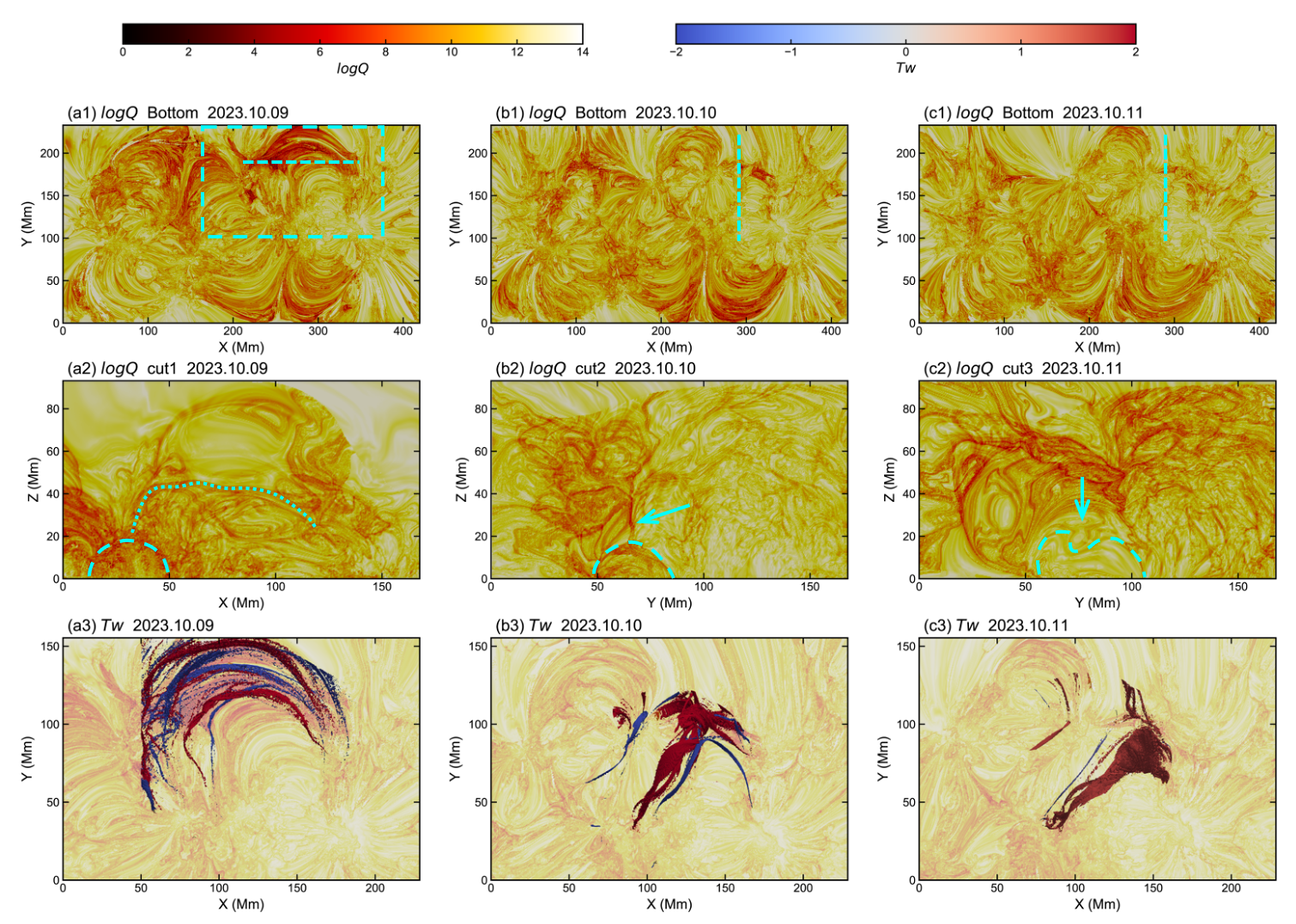}
     \caption{Squashing factor ($Q$) and twist number ($T_w$) derived from the NLFFF models for 2024 October 9--11. The first row shows the photospheric $\log_{10} Q$ maps. The dashed rectangle in panel (a1) outlines the FOV of the third row. The second row shows vertical $\log_{10} Q$ cross-sections taken along the dashed lines marked in the corresponding panels of the first row. The dashed curves in panels (a2) and (b2) outline dome-like QSL structures, and the dotted curve in panel (a2) shows the QSL of the loop system containing the prominence material. The cyan arrow in panel (b2) points to the extended high-$Q$ layer associated with the interaction region. In panel (c2), the cyan arrow and the dashed curve indicate the concave QSL structure where filament material may be located. The third row shows blue--red contours of the projected 3D isosurfaces of enhanced twist, defined by $|T_w| > 2$, overlaid on the corresponding $\log_{10} Q$ maps displayed with an opacity of 0.5.}
     \label{Fig9}
  \end{figure*}

To further examine the magnetic topology associated with the filament reconfiguration, we computed the squashing factor ($Q$) and the twist number ($T_w$) from the NLFFF models \citep{Zhang2022}. The results are shown in Fig.~\ref{Fig9}. On October 9, the photospheric $\log_{10} Q$ map in panel (a1) mainly exhibits a set of layered high-$Q$ footprints associated with the original loop system of AR~13852, indicating that the magnetic connectivity was still dominated by the preexisting AR loops. Near the interface between the emerging region and the filament, the vertical cut in panel (a2) shows that the field lines associated with the observed filament channel are relatively flat and contain shallow dips when viewed from the side (see the dotted curve). A small dome-like quasi-separatrix layer (QSL), whose cross-section is denoted by the dashed curve, is also present near the interaction interface. In panel (a3), the enhanced-$|T_w|$ contours appear as elongated strands aligned with the original filament channel. 

On October 10, after the emergence of AR~13854, the $\log_{10} Q$ distribution changes substantially. An X-shaped high-$Q$ structure appears at the interface between the emerging loops and the preexisting filament-channel field, spatially corresponding to the X-shaped brightening and the candidate reconnection region observed in the SO/EUI-HRI 174~{\AA} images. The vertical $\log_{10} Q$ cuts reveal a dome-like QSL structure, and above it there is a narrow, nearly vertical high-$Q$ layer, as denoted by the cyan arrow in panel (b2). This configuration suggests the formation of a preferential reconnection layer at the interface between the emerging flux and the overlying filament-channel field. At the same time, the enhanced-$|T_w|$ contours become bifurcated, with the branch following the original filament channel becomes fragmented near the X-shaped interaction site, while another branch extends toward the newly forming inter-AR channel. This is consistent with the observed splitting of the filament and the gradual transfer of filament material into the new channel. 

By October 11, the photospheric $\log_{10} Q$ map shows a curved high-$Q$ footprint associated with the newly formed inter-AR filament channel (see panel c1). The vertical cuts in panel (c2) indicate that the initially small dome-like QSL structure has expanded into a larger and more organized QSL system. The strongest high-$Q$ signatures are mainly located along the edge of the dome, especially near the interface between AR~13854 and AR~13852. Inside the dome, the QSL structure shows a concave, low-lying feature (indicated by the cyan curve and arrow), which spatially corresponds to the location of the newly formed filament. This suggests that repeated reconnection has reorganized the emerging-flux domain and the original filament-channel field into a new magnetic connectivity system linking AR~13854 and AR~13852. The enhanced-$|T_w|$ contours are also located along the newly formed filament, supporting the interpretation that the new filament is associated with a non-potential, twisted or strongly sheared magnetic channel. Therefore, the combined $\log_{10} Q$, $T_w$, and H$\alpha$ diagnostics support a scenario in which repeated reconnection transforms the original sheared arcade into a newly formed inter-AR, flux-rope-like or strongly sheared magnetic structure.

\begin{figure}
  \centering
  \includegraphics[trim=0cm 0cm 0cm 0cm, clip, width=\columnwidth]{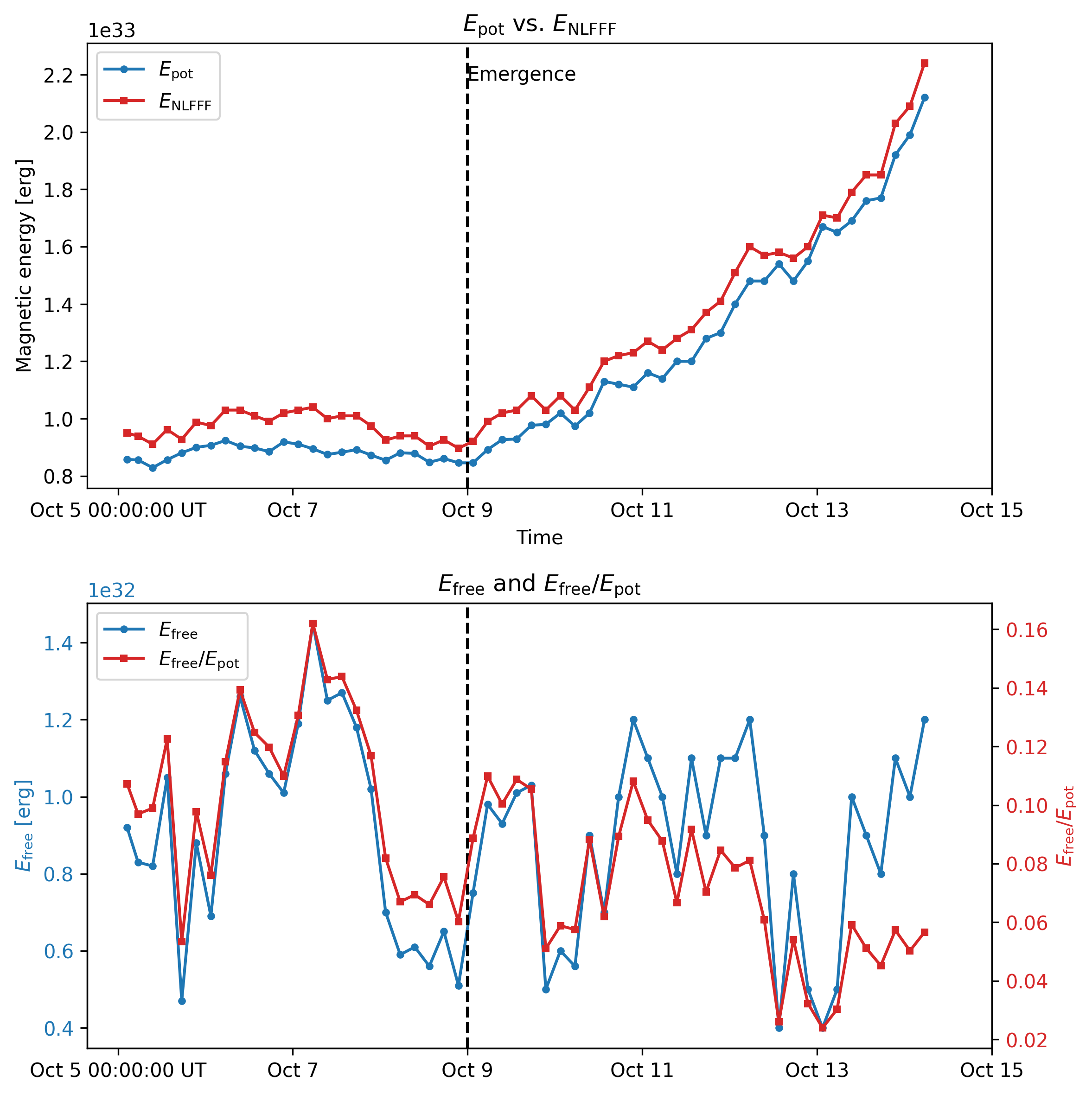}
     \caption{Time evolution of the magnetic energy from the potential-field and NLFFF extrapolations (top), and the corresponding free magnetic energy $E_{\rm free}=E_{\rm NLFFF}-E_{\rm pot}$ together with the ratio $E_{\rm free}/E_{\rm pot}$ (bottom). The vertical dashed line marks the onset of flux emergence of AR~13854 around 2024 October 9.}
        \label{Fig10}
  \end{figure}

Figure~\ref{Fig10} shows the temporal evolution of the magnetic energy derived from the potential-field and NLFFF extrapolations (top panel), together with the free magnetic energy, $E_{\rm free}=E_{\rm NLFFF}-E_{\rm pot}$, and its ratio to the potential-field energy, $E_{\rm free}/E_{\rm pot}$ (bottom panel). The magnetic energy is calculated as $E=(8\pi)^{-1}\int B^2\,dV$ over the whole computational domain. Both the potential and NLFFF energies show little variation before the onset of flux emergence. Around 2024 October 9 (denoted by the dashed line), the total magnetic energy starts to increase rapidly, marking the emergence of AR~13854. As the emergence proceeds, the magnetic energy continues to grow, with the newly emerged active region contributing approximately $1.3\times10^{33}$ erg. 

Unlike the total magnetic energy, the free magnetic energy does not increase monotonically. Instead, it exhibits a variable evolution, with episodes of increase and decrease. The increases are likely associated with the continued injection of shear and electric currents by the emerging AR~13854, whereas the decreases are temporally associated with filament eruptions and flare activity, especially after October 11, indicating the release of stored free energy through reconnection and partial magnetic relaxation. During the evolution, the free energy reaches about 10\% of the potential-field energy, a relatively modest level of non-potentiality. We have examined the robustness of this estimate and find that, although the precise value is subject to some degree of uncertainty inherent to NLFFF extrapolations in general (with additional tests of different optimization iterations indicating the fraction can reach up to about 20\%), this does not affect our overall conclusion that the AR system remains magnetically relatively simple and close to a bipolar configuration throughout the evolution, rather than developing a strongly twisted flux-rope structure of the kind typically associated with more eruptive active regions. This relatively modest free-energy fraction is consistent with the gradual, emergence-driven nature of the filament reconfiguration discussed in this work, rather than the eruption of a preexisting, highly twisted flux rope, and is comparable to free-energy fractions reported in other emerging and moderately active AR systems \citep[e.g.,][]{Aschwanden2013, Wiegelmann2021}.

The formation of the new inter-AR filament does not appear to be controlled simply by a monotonic increase in total magnetic energy. Instead, its timing and morphology are more closely associated with the topological changes revealed by the QSL and $T_w$ diagnostics. In particular, the emergence-driven reconnection rewires the magnetic connectivity between AR~13854 and AR~13852, producing dipped and sheared/twisted field lines capable of supporting the newly formed filament. Thus, while flux emergence supplies magnetic energy and free energy to the system, the filament formation itself is primarily governed by magnetic connectivity reconfiguration in the corona.

\section{Discussion and conclusions}  \label{sec:con}
Using observations that tracked the evolution of two active regions for 11 days, we studied in detail the connectivity changes and filament formation caused by flux emergence. With the combination of large-scale EUV observations, high-resolution H$\alpha$ imaging, and NLFFF modeling, we followed the full sequence from (i) emergence of a second active region adjacent to a preexisting filament channel, through (ii) stepwise rewiring of coronal connectivity, to (iii) the removal of the original filament and the formation of a new inter-AR filament. This sequence is well constrained by the combined photospheric (continuum intensity), chromospheric (H$\alpha$), transition-region (304~{\AA}), and coronal (174~{\AA}) observations.

Our results provide a coherent picture of the filament evolution driven by flux emergence and magnetic reconnection, in which the restructuring proceeds through a spatially and temporally organized sequence of localized reconnection events rather than a single impulsive episode. The early signatures of this interaction are observed in the SST data, where compact emerging bipoles appear near the eastern part of the filament, accompanied by chromospheric brightenings, evolving fibril-like structures, and H$\alpha$ wing signatures of upward and downward motions of filament material. These low-atmosphere signatures indicate that the emerging flux starts to interact with the filament system before the large-scale coronal restructuring becomes fully developed. The subsequent EUV observations reveal that this interaction evolves into a larger-scale magnetic reconfiguration. The observed left-to-right migration of brightenings and bidirectional flows indicates that the magnetic restructuring propagates along the filament channel, forming a reconnection front that progressively modifies the system. 

Our observations suggest that reconnection can operate in a systematic, along-axis mode, consistent with previous reports of propagating brightenings, flows, or activation fronts in filament channels \citep{Yang2019, Li2022}. These results imply that such propagation is not incidental, but may represent an intrinsic mode of magnetic energy release and restructuring in 3D magnetic configurations. A key distinction of the present event is that the magnetic evolution is not confined to a locally perturbed system. Continuous flux emergence alters the large-scale magnetic connectivity, effectively restructuring the filament channel into two independent systems. Consequently, the observed evolution is not a classical eruption–reformation sequence, but rather a topology-driven fragmentation and reorganization process, which also explains the observed change in filament thread orientation. The systematic propagation, together with the asymmetric removal of the filament, suggests that the reconnection front is controlled by the interaction geometry between the emerging AR and the preexisting AR arcades. The longer timescale (hours to days), compared to the rapid evolution in eruptive events, further supports a quasi-static, boundary-driven mechanism instead of an instability-triggered eruption.

The magnetic extrapolations reveal a transition from a predominantly sheared-arcade configuration to a more flux-rope-like configuration. This evolution shares key topological elements with classical reconnection-driven flux-rope formation scenarios \citep{vanBall1989, Moore2001}. In such scenarios, reconnection between the inner legs of sheared arcade fields produces longer, more helical field lines that can contribute to a flux-rope or filament-supporting configuration, while shorter reconnected loops may remain low in the atmosphere and, where appropriate, submerge as flux cancellation. The present event is not a direct realization of these idealized pictures. Here, the reconnection is driven by the emergence of AR~13854 into an already complex coronal environment containing a preexisting filament system associated with AR~13852. The emerging field continuously modifies the large-scale connectivity over an extended region and facilitates repeated reconnection with the preexisting arcade field. Thus, flux emergence acts as the primary driver of the topological reorganization, leading to the removal of the original filament-supporting configuration and the establishment of a new inter-AR filament channel. This interpretation is consistent with modern views of AR formation and coronal restructuring, in which flux emergence, reconnection, and subsequent topological relaxation are tightly coupled processes \citep{Cheung2014}. It is also consistent with recent studies showing that emerging flux can contribute to filament-channel formation through repeated reconnection and plasma injection \citep[e.g.,][]{Wang2019, Li2025}.

A key implication of this event is that filament morphology evolution can be understood as a dynamic process of dip removal and dip formation. The original filament is supported by shallow dips within a sheared arcade, which are progressively removed as the northern branch erupts. At the same time, new dips form in the inter-AR region as the magnetic field evolves toward a more twisted configuration. This interpretation provides a natural explanation for the observed change in filament thread orientation and is consistent with the general understanding that filament plasma accumulates in concave-upward segments of magnetic field lines \citep{Heinzel1999, Mackay2010, Gibson2018}. Importantly, our results extend this picture by demonstrating that dip locations are not static, but can migrate and re-form in response to evolving magnetic connectivity.

Recent modeling work has emphasized that filament fine structure is shaped not only by the large-scale magnetic topology, but also by thermal nonequilibrium, mass drainage and Rayleigh–Taylor–type structuring \citep{Rony2025, Zhou2025b, Valeriia2025}. Simulations by \citet{Zhou2020} showed that apparent counterstreaming signatures may arise from a combination of field-aligned motions and oscillatory behavior, while \citet{Jenkins2022} demonstrated that prominence fine structure and the apparent discrepancy between disk and limb morphology can naturally arise from dynamic evolution in a non-potential magnetic system. In a very recent all-atmosphere 3D radiative MHD study, \citet{Zessner2026} argued that filament formation can begin with stochastic chromospheric injection of cool “seed” material, after which coronal condensation builds additional mass. Our observations are consistent with this broader view: the filament threads do not behave as static tracers of an unchanging field, but instead respond to continuous topological and dynamical reorganization. At the same time, our event places stronger emphasis on emergence-driven topological replacement than on thermal formation alone. We do not have spectroscopic constraints sufficient to determine whether part of the mass of the new inter-AR filament formed through in situ condensation after reconnection, or whether it was dominated by redistribution of preexisting cool material. Nevertheless, the observations strongly suggest that the magnetic pathway for supporting filament plasma changed first, and that the filament body subsequently reorganized within the newly established inter-AR channel. Future spectroscopic observations will be required to distinguish between mass transfer, condensation, and drainage in similar events.

To conclude, the present event demonstrates that flux emergence can drive filament formation not only by supplying magnetic flux and free energy, but also by actively rewiring the magnetic connectivity of a preexisting filament system. In contrast to models that emphasize flux cancellation as the dominant mechanism, our observations indicate that flux emergence can play a comparable or even dominant role, particularly in systems where newly emerged flux interacts strongly with preexisting fields. More broadly, this work highlights the importance of magnetic connectivity evolution as a controlling factor in filament dynamics. The filament observed at any given time may represent only a transient state within a continuously evolving magnetic system. In this sense, filament formation, disappearance, and re-formation should be viewed as different manifestations of the same underlying process: the redistribution of plasma within an evolving network of magnetic dips.

Despite the overall consistency between observations and extrapolations, several limitations should be noted. First, the NLFFF assumption may not be fully valid in the lower atmosphere where plasma forces are non-negligible, which can affect the inferred magnetic dips. Second, the combination of SO/PHI-HRT and SDO/HMI SHARP data introduces uncertainties related to instrument cross-calibration and coordinate transformation \citep{Sinjan2023, Sinjan2026}. Finally, the identification of magnetic dips is based on static extrapolations, whereas the actual filament evolution is inherently dynamic. Future work combining time-dependent data-driven simulations \citep{Inoue2016, Jiang2022} and higher-cadence, high-resolution observations will be essential to further constrain the formation and evolution of filament-supporting magnetic structures.

\begin{acknowledgements}
We thank Solar Orbiter and SDO science teams for the valuable data. Solar Orbiter is a space mission of international collaboration between ESA and NASA, operated by ESA. We are grateful to the ESA SOC and MOC teams for their support. The Active Region Long-term SOOP 2024 was a Solar Orbiter Observing campaign led by C. Froment, X. Li, and H. Eklund, with technical support from C. Fischer, A. Walsh, D. Williams, and all the Solar Orbiter Instruments PIs/teams. The German contribution to SO/PHI is funded by the BMWi through DLR and by MPG central funds. The Spanish contribution is funded by AEI/MCIN/10.13039/501100011033/ and European Union ``NextGenerationEU/PRT'' (RTI2018-096886-C5, PID2021-125325OB-C5, PCI2022-135009-2, PCI2022-135029-2) and ERDF ``A way of making Europe''; ``Center of Excellence Severo Ocho'' awards to IAA-CSIC (SEV-2017-0709, CEX2021-001131-S); and a Ramón y Cajal fellowship awarded to DOS. The French contribution is funded by CNES. This project has received funding from the European Research Council (ERC) under the European Union's Horizon 2020 research and innovation programme (grant agreement No. 101097844 — project WINSUN). TW acknowledges DLR grant 50OC2301. This research is supported by the Research Council of Norway, project number 358540, and through its Centres of Excellence scheme, project number 262622. CF and NP were supported by the Agence Nationale de la Recherche (ANR) for the CROSSWIND project under the grant ANR-24-CE31-2993, the Centre National d’Études Spatiales (CNES), France (ROR: \url{https://ror.org/04h1h0y33}), within the framework of the Solar Orbiter mission, and the Action Thématique Soleil-Terre (ATST) of CNRS/INSU PN Astro, co-funded by CNES and CEA. The Swedish 1-m Solar Telescope (SST) is operated on the island of La Palma by the Institute for Solar Physics of Stockholm University in the Spanish Observatorio del Roque de los Muchachos of the Instituto de Astrof{\'\i}sica de Canarias. The SST is co-funded by the Swedish Research Council as a national research infrastructure (registration number 4.3-2021-00169). We thank the referee for valuable suggestions. 

\end{acknowledgements}

\bibliographystyle{aa} 
\bibliography{aa} 

\appendix
\onecolumn
\section{Validation of the combined SO/PHI-HRT + SHARP boundary condition}
\label{app:sharp_comparison}

To validate the procedure used to construct the extended photospheric boundary condition by combining SO/PHI-HRT and SDO/HMI SHARP vector magnetograms (Section~\ref{sec:res}), we performed potential-field extrapolations using two different boundary conditions: (i) SHARP data alone, and (ii) the combined SO/PHI-HRT + SHARP mosaic. The SO/PHI-HRT data were downsampled to match the native SHARP pixel scale, in order to isolate differences arising from the data-combination procedure itself rather than from differences in spatial resolution.

\begin{figure*}[!htbp]
  \centering
  \includegraphics[width=\textwidth]{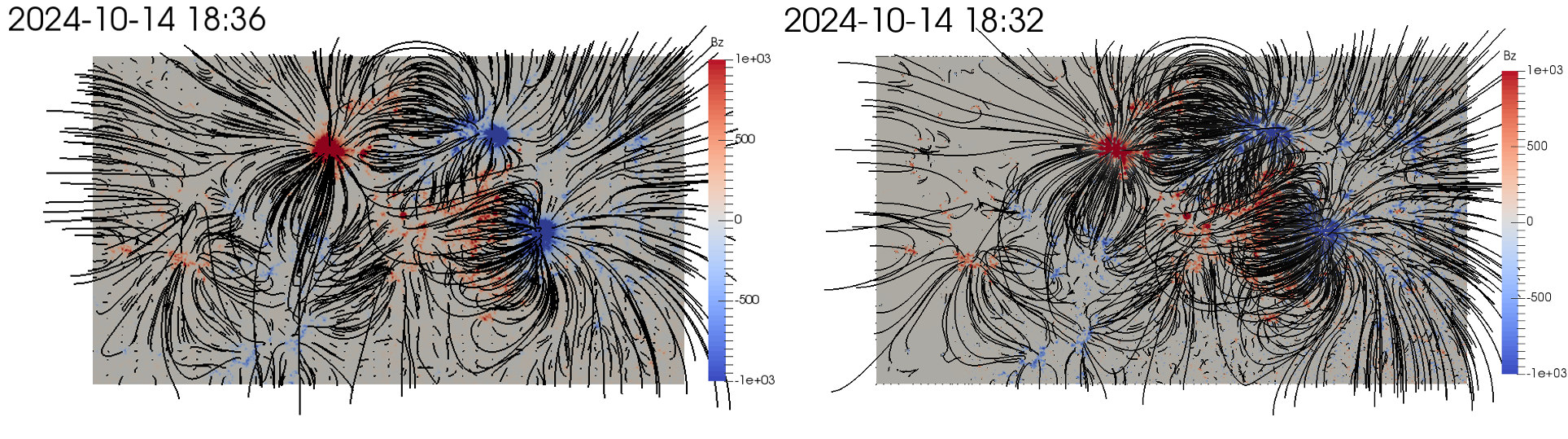}
     \caption{Comparison of potential-field extrapolations using SHARP magnetograms alone (left column) and the combined SO/PHI-HRT + SHARP mosaic (right column) as the photospheric boundary condition on 2024 October 14.}
        \label{FigA1}
\end{figure*}

Figure~\ref{FigA1} compares the resulting extrapolated field lines at representative times. The large-scale magnetic connectivity is broadly consistent between the two cases, with only minor differences that we attribute primarily to the data-processing steps involved in combining the two datasets (e.g., disambiguation and coordinate remapping) and to the 300~G threshold applied to the vector magnetograms (Section~\ref{sec:res}). In particular, this threshold removes part of the weak, quiet-Sun magnetic field outside the AR, resulting in the disappearance of some small-scale loop structures in quiet-Sun regions in the extrapolated field. Within the AR, however, the field lines extrapolated from the downsampled HRT boundary remain well defined and coherent, likely reflecting the higher intrinsic data quality of SO/PHI-HRT even at the degraded (SHARP-matched) resolution. We also note that although the data quality of the SO/PHI-HRT observations decreases somewhat during the later part of the campaign (owing to the increasing spacecraft distance and the associated reduction in signal-to-noise ratio), this has only a minor effect on the extrapolated large-scale active-region structure. These results confirm that the combined HRT+SHARP boundary condition provides a reliable means of extending the temporal and spatial coverage of the magnetic field analysis beyond the field of view and observing cadence of a single instrument alone, while preserving, and in some respects improving, the fidelity of the active-region magnetic structure. We suggest that using SHARP data as the boundary condition for the extended, lower-resolution background field, while employing HRT data to resolve finer structures within its more limited field of view, may represent a practical general strategy for future studies combining Solar Orbiter and SDO data.

\end{document}